\documentclass{article}
\usepackage[english]{babel}
\usepackage{amsmath,amssymb,tabularx,bm,array,booktabs,authblk}
\usepackage{xcolor,slashed,epsfig,cite,float}
\usepackage{verbatim,slashed,ulem,geometry,setspace,algorithm}
\usepackage{algorithmic,mathrsfs,braket,multirow,dcolumn,subfigure,caption,caption}
\usepackage{graphicx,tabulary}
\usepackage{threeparttable}
\graphicspath{{./Figures/}}
\usepackage[colorlinks=true,citecolor=blue,urlcolor=cyan,filecolor=magenta,linkcolor=red,frenchlinks=true,bookmarks]{hyperref}
\allowdisplaybreaks[4]
\renewcommand\sout{\bgroup \color{red}  \ULdepth=-.5ex \ULset}

\begin{document}

\title{Form factors of $\Delta(1232)$ and the electromagnetic $N-\Delta$ transition}

\newcommand{\email}[1]{\thanks{\href{mailto:#1}{#1}}}

\author[a,b]{Jiaqi Wang\email{jqwang@ihep.ac.cn}}
\author[c]{Dongyan Fu\email{fudongyan@impcas.ac.cn}}
\author[a,b]{Yubing Dong\email{dongyb@ihep.ac.cn}}

\affil[a]{Institute of High Energy Physics, Chinese Academy of Sciences, 
Beijing 100049, China}
\affil[b]{School of Physical Sciences, University of Chinese Academy 
of Sciences, Beijing 101408, China}
\affil[c]{Southern Center for Nuclear-Science Theory (SCNT),
\authorcr
Institute of Modern Physics, Chinese Academy of Sciences, Huizhou 516000, China}

\maketitle
\begin{abstract}
In this work, the electromagnetic and gravitational form factors of $\Delta$ isobars, as well as the electromagnetic $N-\Delta$ transition form factors are studied systematically and continuously using a covariant quark-diquark approach with the pion cloud effect.
In our model, the baryon is treated as the two-body system to simplify calculations, and the quarks are assumed to be surrounded by the pion cloud.
The related physical properties, such as the charge radius and magnetic moment of $\Delta$ and the helicity amplitudes of the $N-\Delta$ transition are obtained and discussed.
Our results for the form factors of both $\Delta(1232)$ and the electromagnetic $N-\Delta$ transition are in reasonable agreement with the experimental or lattice results.
Moreover, we found that the pion cloud plays an important role in the results through enlarging the magnetic transition form factor $G_M(t)$ and shifting the sign of the D-term of $\Delta$ to negative.
 
\end{abstract}

\section{Introduction}
The $\Delta (1232)$, as the lowest excited state of the nucleon, is a typical target to study the excited baryon and the strong interaction. However, due to its short lifetime~\cite{ParticleDataGroup:2024cfk}, it is challenging to study its structure directly through experiments. Only the magnetic moments of $\Delta^+$ and $\Delta^{++}$ are respectively obtained through the $\gamma p \rightarrow \pi^0 \gamma' p$ and $\pi^+ p \rightarrow \pi^+ p \gamma$ bremsstrahlung with large uncertainties~\cite{Kotulla:2002cg,LopezCastro:2000cv}. Therefore, theoretical studies about the form factors could serve as a useful tool to study its inner structure and to provide a preliminary prediction to the future experiments.

The electromagnetic form factors (EMFFs) of $\Delta$ have been studied with different models for decades~\cite{Alexandrou:2008bn,Boinepalli:2009sq,Ledwig:2011cx,Li:2016ezv,Kim:2019gka,Sanchis-Alepuz:2013iia,Nicmorus:2010sd}, while the gravitational form factors (GFFs) defined through the matrix element of the symmetric energy-momentum tensor (EMT)~\cite{Pagels:1966zza} have just become a new research topic in recent years.
The mechanical properties, such as the mass radius, the energy and angular-momentum densities, and the inner force distributions, can be obtained through the GFFs~\cite{Polyakov:2018zvc}.
Although it is impractical to measure the GFFs through the graviton scattering, the generalized parton distributions (GPDs) provides an indirect way to measure the GFFs experimentally through the processes such as the deeply virtual Compton scattering (DVCS)~\cite{Diehl:2003ny,Fu:2022bpf}.
Theoretically, the GFFs of $\Delta$ have been studied with different models, including the SU(2) Skyrme model~\cite{Kim:2020lrs}, the covariant quark-diquark model~\cite{Wang:2023bjp,PhysRevD.105.096002}, the QCD sum rule~\cite{Dehghan:2023ytx}, the chiral effective field theory (EFT)~\cite{Alharazin:2022wjj}, and the lattice QCD~\cite{Pefkou:2021fni}.
 
The $N-\Delta$ transition, as another useful tool to study the $\Delta$ resonance, has received lots of attention. The $N-\Delta$ transition has been studied experimentally for decades by different experimental facilities such as CLAS at Jefferson Lab (JLab)~\cite{CLAS:2009ces,CLAS:2001cbm,Mokeev:2022xfo}, A1 and A2 Collaborations at MAMI~\cite{GDH:2004ydy,A1:2008ocu}, LEGS at Brookhaven National Laboratory~\cite{Blanpied:2001ae,Blanpied:1997zz}, and Bates at MIT~\cite{Mertz:1999hp}. Besides, there are lots of theoretical studies with different approaches including the non-relativistic quark model~\cite{Isgur:1981yz}, the relativistic quark model~\cite{Dong:2001js}, the chiral bag model~\cite{Lu:1996rj,Fiolhais:1996bp,Bermuth:1988ms}, the Skyrme model~\cite{Walliser:1996ps}, the quark-diquark model~\cite{Keiner:1996ch}, the chiral EFT~\cite{Pascalutsa:2005ts}, the Sato-Lee model~\cite{Julia-Diaz:2006ios}, the lattice QCD~\cite{Alexandrou:2007dt}, the QCD sum rules~\cite{Wang:2009ru}, the point-form of relativistic quantum mechanics~\cite{Dong:2014gna}, the Dyson-Schwinger approach~\cite{Eichmann:2011aa,Segovia:2013rca}, the chiral quark-soliton model~\cite{Kim:2020lgp}, and so on.
    
In our previous work~\cite{Wang:2023bjp}, we presented a detailed study about the decuplet baryons with a covariant quark-diquark model, where $\Delta$ is treated as a combination of a quark and a diquark to simplify the three-body problem.
Although most of the results are qualitatively consistent with those from other studies, the D-term, a mechanical property derived from the GFFs, has a positive sign and is different from our expectation. As discussed in Ref.~\cite{Perevalova:2016dln}, the D-term is assumed to be negative to guarantee the mechanical stability of the system. In our later study about the nucleon~\cite{Wang:2024abv}, we introduced the pion cloud correction based on Ref.~\cite{Cloet:2014rja}. It is observed that the pion cloud could alter the sign of the D-term into negative. Therefore, in this work, we will present a continuous study
%about the form factors of both $\Delta$ and the $N-\Delta$ transition 
to evaluate the pion cloud effect on $\Delta$ and discuss the electromagnetic structure of the $N-\Delta$ transition.
    
This paper is organized as follows.
Section II gives a brief introduction about the $\Delta$ form factors and the transition form factors.
The quark-diquark approach and the pion cloud effect are also introduced in this section. 
In Sec. III, the parameters employed in this study and the results are presented and discussed. Besides, the pion cloud effect is specially explained.
Finally, a short summary and discussion are given in Sec. IV.

\section{Form Factors And Quark-Diquark Approach}\label{sectionapproach}

\subsection{Form factors of $\Delta$}

For a spin-3/2 particle, the EMFFs are defined through the matrix element of the electromagnetic current, which can be parameterized as~\cite{Cotogno_2020}
\begin{equation}
    \label{ecur}
    \begin{split}
    	\left\langle p',\lambda' \left| \hat{J}^{\mu}\left(0\right) \right|p,\lambda \right \rangle =& -\bar{u}_{\alpha'}\left( p',\lambda' \right)\biggl[ \frac{P^\mu}{M_\Delta} \left( g^{\alpha' \alpha } F^V_{1,0}\left( t \right) -\frac{q ^{\alpha'} q ^\alpha}{2M_\Delta^2} F^V_{1,1}\left( t \right)\right)\\
        &+\frac{i \sigma^{\mu q}}{2M_\Delta} \left( g^{\alpha' \alpha} F^V_{2,0}\left( t \right)-\frac{q ^{\alpha'} q ^\alpha}{2M_\Delta^2}F^V_{2,1}\left( t \right)\right) \biggr]  u_\alpha \left( p,\lambda \right),
    \end{split}
\end{equation}
where $i \sigma^{\mu q}=i \sigma^{\mu\rho}q_\rho$, $M_\Delta$ stands for the $\Delta$ mass and $u_\alpha \left( p,\lambda\right)$ is the Rarita-Schwinger spinor with the normalization relation as $\bar{u}_{\sigma'}(p) u_\sigma(p)=-2 M_\Delta \delta_{\sigma' \sigma}$. The kinematical variables introduced in Eq.~\eqref{ecur} are defined as $P^\mu=\left(p^\mu+p'^\mu\right)/2$, $q^\mu=p'^\mu-p^\mu$, and $t=q^2$, where $p$ ($p'$) is the initial (final) momentum. It should be mentioned that the form factors are contributed by both the quark and the gluon. In this work, we only consider the constituent quark contribution.
    
The EMFFs, including the electric-monopole (-quadrupole) form factor $G_{E0}$ ($G_{E2}$) and the magnetic-dipole (-octupole) form factor $G_{M1}$ ($G_{M3}$), are defined through the linear combinations of $F^V_{i,j}$. 
The detailed definitions of the EMFFs can be found in our previous work \cite{PhysRevD.105.096002}, and we do not repeat them here. When the squared transfer momentum $t$ goes to zero, the electric charge $Q_e$, magnetic moment $\mu$, electric-quadrupole moment $\mathcal{Q}$, and magnetic-octupole moment $\mathcal{O}$ can be obtained through the relations $Q_e=G_{E0}(0)$, $\mu=\frac{e}{2M_\Delta} G_{M1}(0)$, $\mathcal{Q}=\frac{e}{M_\Delta^2} G_{E2}(0)$, $\mathcal{O}=\frac{e}{2M_\Delta^3} G_{M3}(0)$~\cite{RAMALHO2009355}. Moreover, the electric charge and magnetic radii are defined from their corresponding form factors as~\cite{Leinweber:1992hy}
\begin{equation}
    {\langle r^2\rangle}_{E0}=\left.\frac{6}{G_{E0}(0)}\frac{d}{dt}G_{E0}(t)\right|_{t=0},\quad {\langle r^2\rangle}_{M1}=\left.\frac{6}{G_{M1}(0)} \frac{d}{dt}G_{M1}(t)\right|_{t=0} \footnote{For $\Delta^0$, the charge radius is defined as~\cite{Li:2016ezv}
    \[\langle r^2\rangle_{E0}=\left.6 \frac{d}{dt}G_{E0}(t)\right|_{t=0}.\nonumber\]}.
\end{equation}

Similarly, the GFFs are defined through the matrix element of the symmetric energy-momentum tensor $\hat{T}^{\mu \nu}$, which reads~\cite{Kim:2020lrs}
\begin{equation}
    \label{emtcur}
    \begin{split}
        &\left\langle p^\prime,\lambda^\prime \left| 
        \hat{T}^{\mu \nu}(0)\right| p,\lambda\right\rangle  \\
        =&-\bar{u}_{\alpha ^\prime}\left(p^\prime,\lambda^\prime\right) 
        \bigg [\frac{P^\mu P^\nu}{M_{\Delta}}\left(g^{\alpha'\alpha} 
        F^T_{1,0} (t)-\frac{q ^{\alpha' } 
        q ^{\alpha}}{2 M_\Delta^2}F^T_{1,1} (t) \right)
        + \frac{ \left({q }^\mu {q }^\nu- {g}^{\mu \nu}q^2\right)}{4M_\Delta}    
        \left({g}^{\alpha'\alpha}F^T_{2,0} (t)
        -\frac{{q }^{\alpha'}{q}^{\alpha}}{2 M_\Delta^2}F^T_{2,1} (t)\right)\\
        &+ M_\Delta g^{\mu  \nu} \left(g^{\alpha'\alpha}F^T_{3,0}(t)
        -\frac{ q^{\alpha'} q^{\alpha}}{2M_\Delta^2}F^T_{3,1} (t)\right)
        + \frac{i {P}^{ \{ \mu } \sigma ^{\nu \} \rho}q_\rho}{2M_\Delta} 
        \left(g^{\alpha' \alpha}F^T_{4,0} (t) 
        -\frac{q ^{\alpha'} q ^{\alpha}}{2 M_\Delta^2}F^T_{4,1}(t)\right) \\
        &- \frac{1}{M_\Delta} \left({q }^{\{ \mu} g^{\nu \} 
        \{ \alpha '} {q }^{\alpha \}}-2 q^{\alpha' }q^{\alpha} g^{\mu \nu }
    - g^{\alpha' \{ \mu } g^{\nu \} \alpha } q^2 \right) F^T_{5,0} (t)    
    + M_\Delta g^{\alpha ' \{ \mu } g^{\nu \} \alpha}F^T_{6,0}(t) 
    \bigg]u_\alpha \left(p,\lambda\right),
    \end{split}
\end{equation}
where the convention $a^{\{\mu}b^{\nu\}}=a^\mu b^\nu+a^\nu b^\mu$ is used. 
Notice that $F^T_{3,0}$, $F^T_{3,1}$ and $F^T_{6,0}$ are the non-conserving terms which will vanish for the conserving current.
% when considering the contribution from the gluon. 
% Therefore, we simply ignore them here.
The GFFs, including the energy-monopole (-quadrupole) form factor ${\varepsilon}_{0(2)}$, the angular momentum-dipole (-octupole) form factor $\mathcal{J}_{1(3)}$, and $D_{0,2,3}$, which are connected with the pressure and shear force distributions inside the particle analogizing the classical physics\cite{Polyakov:2018zvc}, can be expressed as the linear combinations of $F^T_{i,j}$~\cite{Kim:2020lrs}.
    
The mechanical preperties of $\Delta$ can be derived from the GFFs. The definitions of the mass and mechanical radii are
\begin{equation}
\label{MassRadius}
    \langle r^2 \rangle_\text{m} = \left.\frac{6}{\varepsilon_0(0)} \frac{d}{dt}\varepsilon_0(t) \right|_{t=0},\qquad
    \langle r^2 \rangle_\text{mech} =\frac{6D_0(0)}{\int^0_{-\infty}dt D_0(t)}.
\end{equation}

Through the Fourier transformation to the coordinate space, one gets the energy-monopole (-quadrupole) density $\mathcal{E}_{0(2)}(r)$ and the angular momentum density $\rho_J(r)$ of $\Delta$.
The pressure distribution $p_0(r)$, the shear force distribution $s_0(r)$, and their high-order distributions $p_{2,3}(r)$ and $s_{2,3}(r)$, are also obtained through the Fourier transformation from their corresponding form factors~\cite{Kim:2020lrs}. 
    
\subsection{$N-\Delta$ transition form factors}

The matrix element of the $N-\Delta$ transition current is decomposed as~\cite{Pascalutsa:2006up}
\begin{equation}
    \begin{split}
        \left\langle \Delta(p',\lambda') \left| \hat{J}^{\mu}\left(0\right) \right| N(p,\lambda) \right \rangle  =& \sqrt{\frac{3}{2}} \frac{M_{\Delta} + M_N}{M_N\left[(M_{\Delta} + M_N)^2-t\right]} \bar{u}_{\alpha'}(p',\lambda') \left [ g_M(t) i \epsilon^{\alpha' \mu p' q} \right. \\
        &\left. - g_E(t) \left( q^{\alpha'} p'^{\mu} - q \cdot p' g^{\alpha' \mu} \right) \gamma_5  - g_C(t) \left( q^{\alpha'} q^{\mu} - q^2 g^{\alpha' \mu} \right) \gamma_5 \right] u(p,\lambda),
    \end{split}
\end{equation}
where $\epsilon^{\alpha' \mu p' q}=\epsilon^{\alpha' \mu \lambda \rho} p'_{\lambda} q_{\rho}$, $M_N$ is the nucleon mass, $u\left(p,\lambda\right)$ is the Dirac spinor with normalization as $\overline{u}\left(p,\lambda\right)u\left(p,\lambda\right)=2M_N$, and the convention $\epsilon_{0123}=+1$ is used. 
The magnetic-dipole (M1) form factor $G_M$, electric-quadrupole (E2) form factor $G_E$ and Coulomb-quadrupole (C2) form factor $G_C$ are defined as follows, 
\begin{subequations}
    \begin{align}
        G_M(t) &= g_M(t) + \frac{1}{Q_+^2} \left[ \frac{1}{2} (-M_\Delta^2 + M_N^2 -t) \, g_E(t) -t g_C(t)  \right], \\
        G_E(t) &= \frac{1}{Q_+^2} \left[ \frac{1}{2} (-M_\Delta^2 + M_N^2-t) \, g_E -t  g_C(t)\right], \\
        G_C(t) &= \frac{1}{Q_+^2} \left[ (-M_\Delta^2 + M_N^2-t) \, g_C(t) - 2M_\Delta^2 \, g_E(t) \right],
    \end{align}
\end{subequations}
where $Q^2_\pm=(M_\Delta \pm M_N)^2-t$. 

The transition magnetic moment $\mu_{N-\Delta}$ and electric-quadrupole moment $\mathcal{Q}_{N-\Delta}$ are derived from the corresponding form factors with $t=0$ as
\begin{equation}
    \mu_{N - \Delta}=\sqrt{\frac{M_\Delta}{M_N}}\frac{e}{2M_N}G_M(0),
\end{equation}
\begin{equation}
    \mathcal{Q}_{N -\Delta}=-6\sqrt{\frac{M_\Delta}{M_N}}\frac{2e M_\Delta }{M_N(M_\Delta^2-M_N^2)}G_E(0).
\end{equation}  
    
The quadrupole form factors can be presented as the multipole ratios $R_{EM}(t)$ and $R_{SM}(t)$, also known as EMR and CMR, which are defined as
\begin{equation}
    R_{EM}(t)=-\frac{G_E(t)}{G_M(t)}, \quad R_{SM}(t)=-\frac{Q_+ Q_-}{4M_\Delta^2}\frac{G_C(t)}{G_M(t)}.
\end{equation}
Besides, the helicity amplitude is another observables extracted through the experiments directly.
The relations between the helicity amplitudes and the transition form factors are
\begin{subequations}
    \begin{align}
        A_{3/2}(t) &= -N \frac{\sqrt{3}}{2} \left( G_M(t) + G_E(t) \right), \\
        A_{1/2}(t) &= -N \frac{1}{2} \left( G_M(t) - 3G_E(t) \right), \\
        S_{1/2}(t) &= N \frac{Q_+ Q_-}{2 \sqrt{2} M_\Delta^2} G_C(t),
    \end{align}
\end{subequations}
where $N=\frac{e}{2} \left( \frac{Q_+ Q_-}{2M_N^3} \right)^{1/2} \frac{M_N + M_\Delta}{Q_+}$. 

\subsection{Quark-diquark approach}

In the quark-diquark approach, the baryon is treated as a combination of the quark and the diquark. 
Based on the flavour wave functions listed in the~\ref{appendix}, the total matrix element is expressed as the sum of the quark and diquark contributions. 
The detailed calculation process of $\Delta$ can be found in our previous work~\cite{PhysRevD.105.096002}, and hence we only introduce the model with the electromagnetic $N-\Delta$ transition briefly.

\begin{figure}[htbp]
    \centering
    \includegraphics[width=0.9\linewidth]{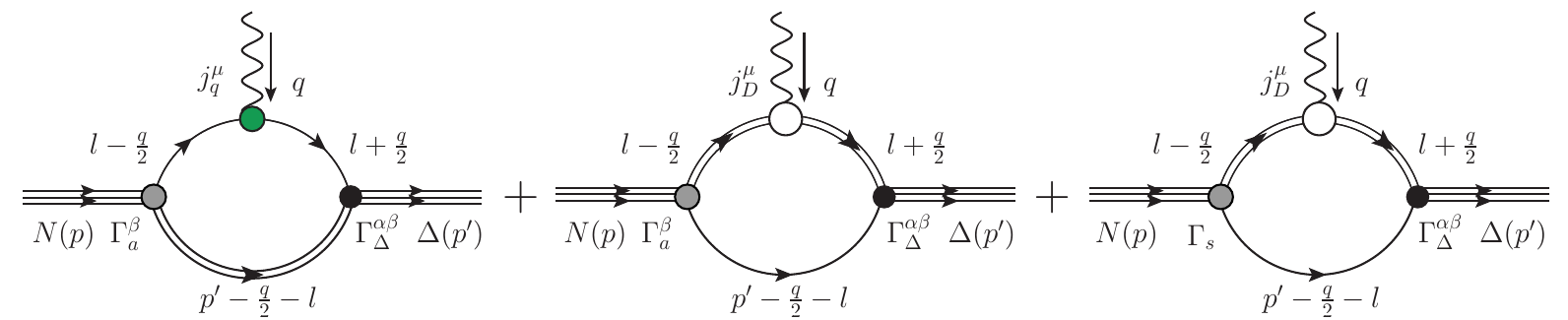}
    \caption{\small{Feynman diagrams for the transition current coupling with the quark (the first panel) and the diquark (the second and third panels). The grey and black points are the baryon-quark-diquark vertices for the nucleon and $\Delta$, respectively, and the green point stands for the dressed photon-quark vertex introduced in Sec. \ref{PionCloud}. The internal structure of the diquark is considered by coupling with the photon (the white points) and shown later in Fig.~\ref{f-dq}.}}
    \label{f-emff}
\end{figure}

Since the nucleon is a spin-1/2 particle, the spin of the diquark inside can be spin-0 (scalar) or spin-1 (axialvector).
Therefore, the transition form factors are contributed by three parts, the quark, the axialvector diquark and the scalar-axialvector diquark transitions, corresponding to the three diagrams in Fig.~\ref{f-emff}, where $\Gamma^\beta_{s(a)}$ and $\Gamma^{\alpha \beta}_\Delta$ are the baryon-quark-diquark vertices for the nucleon and $\Delta$, respectively, and the subscript $s$ or $a$ represents that the diquark inside the nucleon is scalar or axialvector. 

Based on Fig.~\ref{f-emff}, one writes down the quark contribution as
\begin{equation}
    \label{NDmatrix1}
    \begin{split}
        & \left\langle \Delta (p^\prime,\lambda^\prime) \left| \hat{J}^{\mu}_{q}(0) \right| N( p,\lambda)\right\rangle    \\
        = & i  \bar{u}_{\alpha'}(p',\lambda') \int \frac{d^4 l}{(2 \pi)^4}\frac{1}{\mathfrak{D}} \widetilde{\Gamma}^{\alpha' \beta'}_\Delta \left( 
        \slashed{l}+\frac{\slashed{q}}{2}+m_q \right) (-g_{\beta' 
        \beta}) \Lambda^{\mu}(t) \left( \slashed{l}-\frac{\slashed{q}}{2} +m_q 
        \right)\widetilde{\Gamma}^{ \beta}_{a} u(p,\lambda),
    \end{split}
\end{equation} 
where $\mathfrak{D}$ contains all the denominators in the propagators.
It should be noted that we neglect the momentum related term in the axialvector diquark propagator to avoid the integral divergency and simplify the calculation.
In our previous work about the nucleon~\cite{Wang:2024abv}, we have analyzed the effect of this term, and it reveals that the contribution is nearly negligible. The modified vertex $\widetilde{\Gamma}$ in eq. (11) is the baryon-quark-diquark vertex combined with an additional scalar function $\Xi(p_1,p_2)$ as $\widetilde{\Gamma}=\Xi(p_1,p_2)\Gamma$ to ensure that the quark and the diquark can form a bound state. The baryon-quark-diquark vertices are borrowed from Ref.~\cite{Scadron_1968} as 
\begin{subequations}
    \begin{align}       
        \Gamma_{s}&=c_s,\\ 
        \Gamma_{a}^\beta &=c_a \gamma^\beta \gamma^5,\\
        \Gamma^{\alpha\beta}_\Delta&=c_\Delta\left(g^{\alpha\beta}+c_1 p_r^\alpha\gamma^\beta  +c_2 p_r^\alpha p_r^\beta\right), \label{vertexfunction}     
    \end{align}
\end{subequations}
where $p_r$ is the relative momentum between the quark and the diquark, $c_1$, $c_2$ in Eq.~\eqref{vertexfunction} can be determined by fitting to the lattice data~\cite{Alexandrou:2008bn}, and $c_s$, $c_a$ and $c_\Delta$ are normalization constants. The scalar function $\Xi$~\cite{PhysRevD.80.054021} reads
\begin{equation}
\label{vertexfunction2}
    \Xi (p_1,p_2)=\frac{c_0}{\left[ p_1^2 -m_R^2
    +i \epsilon\right]\left[ p_2^2 -m_R^2+i \epsilon\right]},
\end{equation}
where $m_R$ is a cutoff parameter which is positively correlated with the baryon mass, $c_0$ is a coefficient in unit of $\text{GeV}^4$, and $p_1$, $p_2$ are the momentums of the quark and diquark at the vertex. It should be mentioned that the momentum dependent scalar function $\Xi(p_1,p_2)$ may break the gauge invariance and the Ward-Takahashi identity~\cite{Broniowski:2008hx,Davidson:1994uv}. Therefore, to simplify the calculations, the normalization constants $c_s$, $c_a$ and $c_\Delta$ are determined by hands to satisfy $G_{E0}(0)=1$ for $\Delta^+$ and $G_E(0)=1$ for the proton.

\begin{figure}[htbp]
    \centering
    \includegraphics[width=0.3\linewidth]{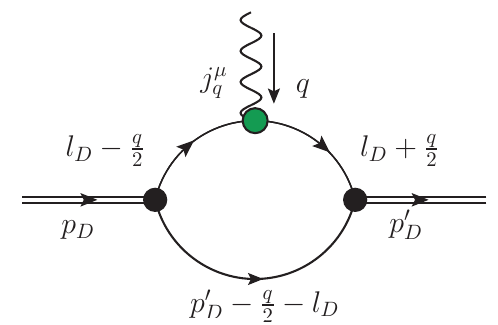}
    \caption{\small{Feynman diagram for the internal structure of the diquark, where the black points are the quark-diquark coupling vertices with $\gamma^5$ for the scalar diquark and $\gamma^\alpha$ for the axialvector diquark. The kinematical variables have the relations $p_D=l-\frac{q}{2}$ and $p'_D=l+\frac{q}{2}$, where $l$ and $q$ are the momentums in the second or third panel of Fig.~\ref{f-emff}.}}
    \label{f-dq}
\end{figure} 
    
When calculating the second and third digrams in Fig.~\ref{f-emff}, the internal structure of the diquark is considered. The structure of the diquark can be described as an effective current. For the axialvector diquark, it reads 
\begin{equation}   
    \label{dqmatrix1}
    j_{D,a}^{\mu,\beta'\beta}= \left[ g^{\beta'\beta} F_{a,1}(t) - \frac{q^{\beta'} q^\beta}{2 m_a^2} F_{a,2}(t) \right] (p_D' + p_D)^\mu - \bigl( q^{\beta'} g^{\mu\beta} - q^\beta g^{\mu\beta'} \bigr) F_{a,3}(t),
\end{equation}
where the form factors $F_{a,1}$, $F_{a,2}$ and $F_{a,3}$ are obtained through the Feynman diagram in Fig.~\ref{f-dq}.
For the scalar-axialvector transition, it reads 
\begin{equation}    
    \label{dqmatrix2}
    j_{D,sa}^{\mu,\beta'}=i \epsilon^{\beta' \mu p_D' q} F_{sa}(t),
\end{equation}
where $F_{sa}$ is obtained through Fig.~\ref{f-dq} with the scalar diquark in the nucleon and axialvector diquark in $\Delta$.

With the effective current, one respectively writes down the matrix elements of the second and third diagrams in Fig.~\ref{f-emff} as
\begin{equation}
    \label{NDmatrix2}
    \begin{split}
        \left\langle \Delta (p^\prime,\lambda^\prime) \left| \hat{J}^{\mu}_{D,a}(0) \right| N( p,\lambda)\right\rangle = & i\bar{u}_{\alpha'}(p',\lambda')\int \frac{d^4 l}{(2 \pi)^4}\frac{1}{\mathfrak{D}}\widetilde{\Gamma}^{\alpha'}_{\Delta, \beta'} \left( 
        \slashed{p}'-\frac{\slashed{q}}{2}-\slashed{l}+m_q \right) j_{D,a}^{\mu,\beta'  \beta}\widetilde{\Gamma}_{a,\beta} u(p,\lambda),
    \end{split}
\end{equation} 
and
\begin{equation}
    \label{NDmatrix3}
    \begin{split}
        \left\langle \Delta (p^\prime,\lambda^\prime) \left| \hat{J}^{\mu}_{D,sa}(0) \right| N( p,\lambda)\right\rangle = & i\bar{u}_{\alpha'}(p',\lambda')\int \frac{d^4 l}{(2 \pi)^4}\frac{1}{\mathfrak{D}}\widetilde{\Gamma}^{\alpha' }_{\Delta, \beta'} \left( 
        \slashed{p}'-\frac{\slashed{q}}{2}-\slashed{l}+m_q \right) j_{D,sa}^{\mu,\beta'}\widetilde{\Gamma}_{s} u(p,\lambda).
    \end{split}
\end{equation}

\subsection{Pion cloud effect}\label{PionCloud}

\begin{figure}[htp]
	\centering
    \includegraphics[width=0.9\linewidth]{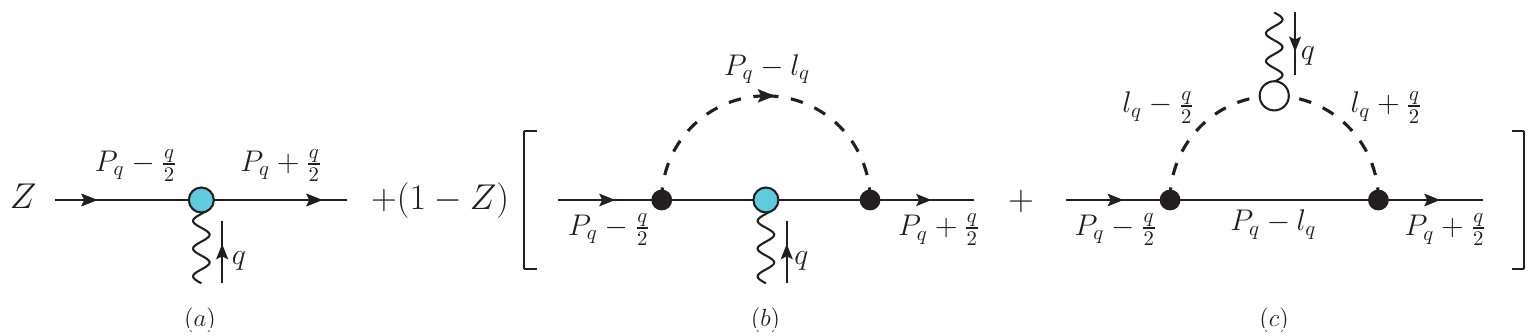}
	\caption{\small{The photon-quark coupling process with the pion cloud correction.
    Diagram (a) shows the coupling process without the pion cloud.
    In diagrams (b) and (c), the electromagnetic current respectively couples with the quark and the pion, and the form factors $f_{i,0}^q$ and $f_{i,0}^{\pi}\, (i=1,2)$ are originated from these two diagrams.
    Specially, when calculating diagrams (b) and (c), the quark is assumed to be point-like to simplify the calculation.
    The blue points stand for the bare vertex $J^{\mu}_{PL}=\gamma^\mu$, and the white point represents the inner structure of the pion.}}
	\label{pcCoupling}
\end{figure}

In this paper, it is assumed that the quark is no longer a point-like particle, but a structured quark surrounded by the pion cloud. As a consequence, the interaction vertex between the quark and the electromagnetic current or the EMT current is modified accordingly, which reads
\begin{equation}
    \label{pqvertex}
        \Lambda^{\mu} \left( t \right)=\gamma^{\mu} F_{1}^q(t)+ \frac{i \sigma^{\mu q}}{2m_q}F_{2}^q(t),
\end{equation}  
for the electromagnetic current and
\begin{equation}
    \label{gqvertex}
        \Lambda^{\mu\nu} \left( t \right)=\frac{1}{2} P_q^{\{\mu} \gamma^{\nu\}} F_{1,0}^q(t)+\frac{\left(q^\mu q^\nu-g^{\mu\nu}q^2\right)}{m_q}F_{2,0}^q(t)+m_q g^{\mu\nu}F^q_{3,0}(t)+\frac{i}{4m_q}P_q^{ \{ \mu}\sigma^{\nu\}q}F_{4,0}^q(t),
\end{equation}  
for the EMT current, where $P_q^\mu=\left(p_q^\mu+p_q'^\mu\right)/2$ is the average momentum between the initial and final quarks, and the form factors $F^q_i$ or $F^q_{i,0}$ are the mixture between the point-like quark and the dressed quark. Taking the electromagnetic coupling vertex as an example, $F^q_{i}(t)$ can be expressed as~\cite{Cloet:2014rja}
\begin{equation}
    \label{gqvertex124}
        F_{i}^q=Z F_{i}^{PL} +\left(1-Z\right)(f_{i}^q +f_{i}^\pi),\quad \text{with} \quad i=1,2,
\end{equation} 
where $F_{1}^{PL}=Q_e$ and $F_{2}^{PL}=0$ are the form factors of the point-like quark and $Q_e$ is the charge quantum number carried by the quark.
The $f_{i}^q$ and $f_{i}^{\pi}$ ($i=1,2$) are the form factors originated from the second and third diagrams in Fig.~\ref{pcCoupling}, respectively.
Moreover, the factor $Z$ stands for the possibility to strike a quark without the pion cloud, which is derived from the self-energy correction of the pion loop~\cite{Cloet:2014rja}.    
 
Similarly, for the EMT vertex, the point-like form factors are defined as $F_{1,0}^{PL}=1$, $F_{i,0}^{PL}=0$ ($i=2,3,4$).
The dressed form factors $f_{i,0}^q$ and $f_{i,0}^{\pi}$ ($i=1,2,3,4$) can be obtained through calculating the GFFs of the dressed quark.
The detailed calculation process can be found in our previous work about the nucleon~\cite{Wang:2024abv}.    

\section{Numerical Results}\label{sectionresults}

\subsection{Parameter determination} \label{ParaD} 

To ensure the bound state, the input masses must satisfy the relations $M_B<m_q+m_D$ and $m_D<2 m_{q}$, where $M_B$ stands for the baryon mass and $m_D$ is the diquark mass.
The $\Delta$ mass $M_\Delta$, the nucleon mass $M_N$ and the pion mass $m_{\pi}$ are taken from Ref.~\cite{ParticleDataGroup:2024cfk}.
In our previous work about the nucleon, we choose $m_q=0.38\,\text{GeV}$ and $m_s=m_a=0.70\,\text{GeV}$. However, this set of parameters can not satisfy the $\Delta$ mass relations mentioned above with $M_\Delta=1.232\,\text{GeV}$. 
Therefore, we choose another set of parameters, $m_a=0.88\,\text{GeV}$, $m_s=0.70\,\text{GeV}$ and $m_q=0.45\,\text{GeV}$.
It should be stressed that we can still obtain reasonable nucleon form factors using the new parameters in this work.
The different parameters leave little effect on the charge radii and magnetic moments of the nucleon, and the normalization conditions of the GFFs can still be satisfied.
The cut-off parameter $m_R$ varies around the mass of the baryon. Here we choose $m_R=1.4 \, \text{GeV}$, larger than that of the nucleon~\cite{Wang:2024abv}.
The $\theta$ introduced in the~\ref{appendix} is a mixing angle characterizing the proportion between the scalar diquark and the axialvector diquark inside the nucleon.
We choose $\text{sin}^2\theta=0.5$, the same with our work about the nucleon~\cite{Wang:2024abv}.
The normalization coefficients $\mathcal{C}_\Delta=c_0 c_\Delta$, $\mathcal{C}_s=c_0 c_s$ and $\mathcal{C}_a=c_0 c_a$ are determined through the normalization condition $G_{E0}(0)=1$ of $\Delta^+$ and the proton.
The couplings $c_1$, $c_2$ in the quark-diquark vertex \eqref{vertexfunction} are determined through comparing with the lattice data in Ref.~\cite{Alexandrou:2008bn}.
All the parameters used in this paper are listed in Tab.~\ref{para}.
\begin{table}[H]
        \renewcommand\arraystretch{1.3}
		\centering
        \caption{\small{Parameters used in this work.}} 

		\begin{tabular}{ccccccc}
     		\toprule
     		\toprule
                $M_{\Delta}$/GeV & $M_N$/GeV & $m_s$/GeV & $m_a$/GeV & $m_q$/GeV & $m_\pi$/GeV  & $Z$  \\
        	\midrule
        	1.232 & 0.938 & 0.70 & 0.88 & 0.45 & 0.14 & 0.73\\
            \midrule
            \midrule
            $\text{sin}^2\theta$ & $\mathcal{C}_\Delta\text{/GeV}^4$ & $\mathcal{C}_s\text{/GeV}^4$ & $\mathcal{C}_a\text{/GeV}^4$ & $c_1/\text{GeV}^{-1}$ &
            $c_2/\text{GeV}^{-2}$ & 
            \\
        	\midrule
        	0.5 & 13.29 & 13.02 & 13.00 & 0.71 & 0.63 & \\
        	\bottomrule
        	\bottomrule
     	\end{tabular}     	
		\label{para}  
    \end{table} 

\subsection{Electromagnetic form factors of $\Delta$}

\begin{figure}[H]
        \centering
        \includegraphics[width=0.46\linewidth]{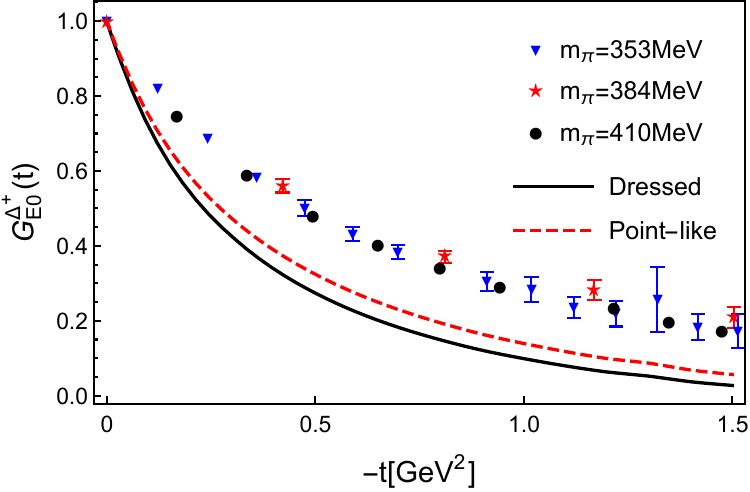}
        \includegraphics[width=0.47\linewidth]{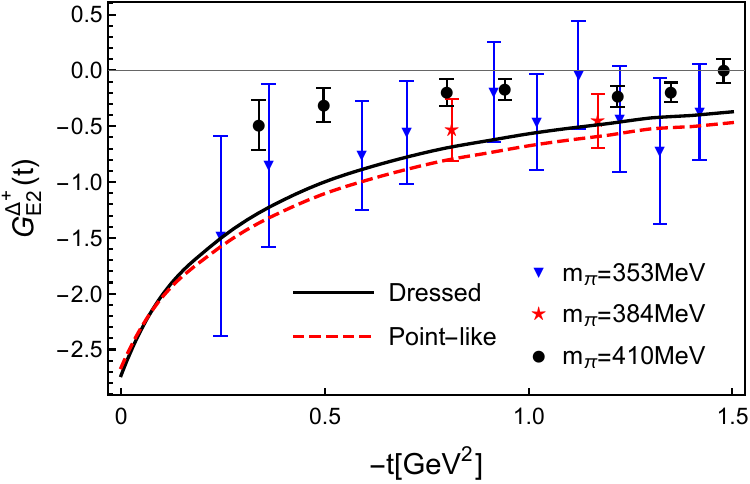}\\
        \, \includegraphics[width=0.45\linewidth]{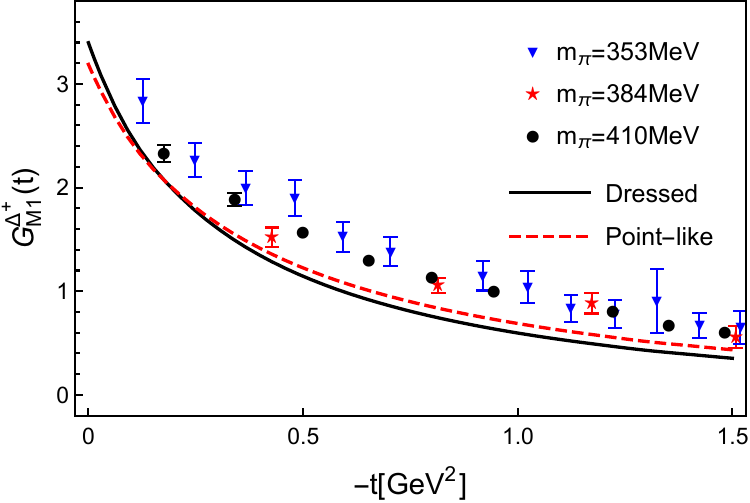}
        \includegraphics[width=0.47\linewidth]{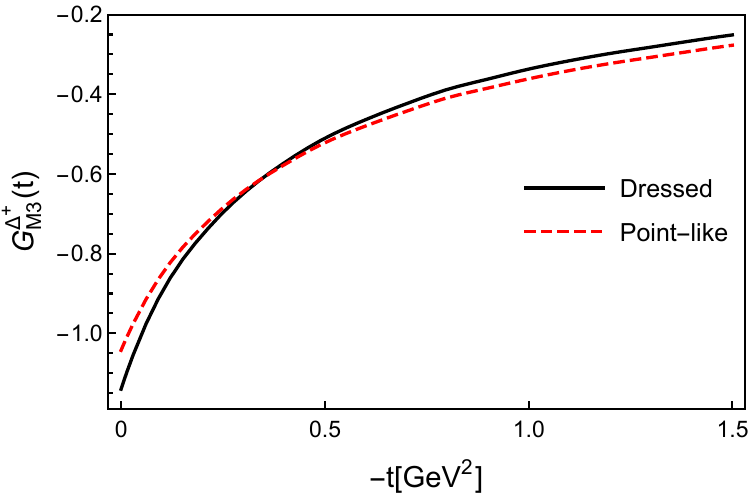}
        \caption{\small{EMFFs of $\Delta^+$.
        The results are compared with the lattice data in Ref.~\cite{Alexandrou:2008bn}, with pion mass $m_{\pi}=353$ MeV (blue triangles), $m_{\pi}=384$ MeV (red stars), and $m_{\pi}=410$ MeV (black points).}}
        \label{EMFigure}
    \end{figure}

The calculated EMFFs of $\Delta^+$ are illustrated in Fig.~\ref{EMFigure} compared with the lattice QCD data~\cite{Alexandrou:2008bn}.
The black solid and red dashed curves in all the result figures of this work respectively represent the results with and without the pion cloud correction.
As seen in Fig.~\ref{EMFigure}, the pion cloud slightly increases the magnitudes of the magnetic moment, electric-quadrupole moment and charge radius of $\Delta^+$, and the results are in qualitatively consistent with the lattice results. 

The electromagnetic properties of $\Delta$ resonances, including the charge radii, magnetic moments, magnetic radii, and electric-quadrupole moments are listed in Tabs.~\ref{ChargeRadius}-\ref{EQM}, compared with the experimental data and other studies. We found that the pion cloud effect enlarges all the magnitudes of the properties, especially for $\Delta^-$.
Without the pion cloud effect, contributions from $u$ and $d$ quarks are assumed to be identical except for their charges, and hence the form factors of different $\Delta$ isobars have strict multiple relations with their charges.
However, the pion cloud effect implies that the corrected photon-quark coupling vertices in Eq.~\eqref{pqvertex} are different to the $u$ and $d$ quarks.
As a result, the form factor $F_2^q(0)$ of the $d$ quark in the effective vertex has larger magnitude than that of the $u$.
Therefore, the pion cloud effect on the $d$ quark is stronger.
Taking the magnetic moments as an example, the pion cloud increases the magnitudes of the magnetic moments of $\Delta^{++}$ and $\Delta^-$ by about $17\%$ and $51\%$. Moreover, the pion cloud effect on $\Delta^+$ is the weakest, because the effects from the $u$ and $d$ quarks are partly cancelled. 

\begin{figure}[H]
        \centering
        \includegraphics[width=0.32\linewidth]{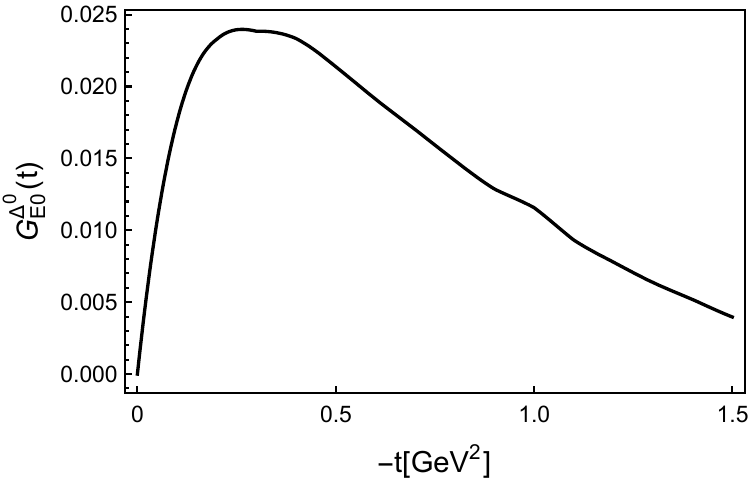}
        \includegraphics[width=0.32\linewidth]{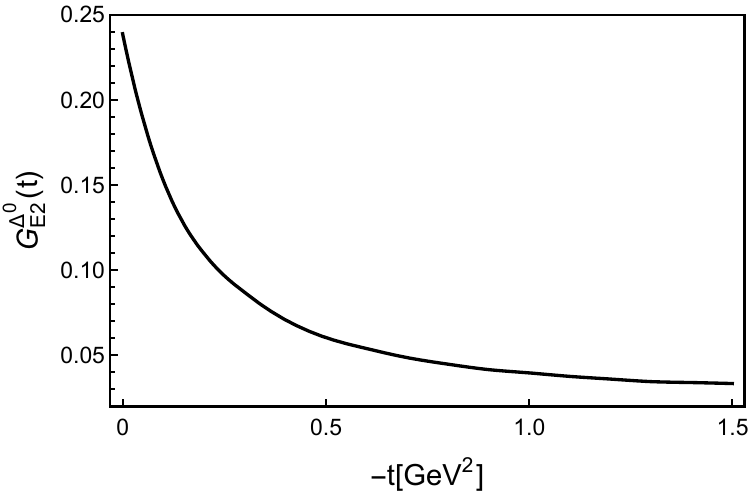}
        \includegraphics[width=0.32\linewidth]{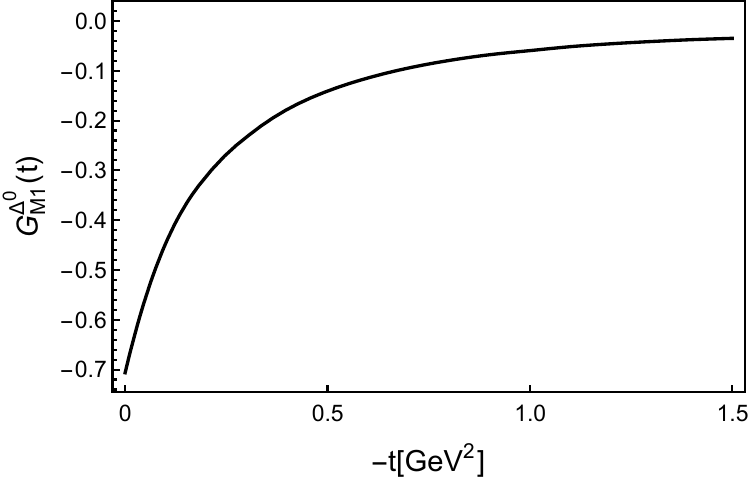}
        \caption{\small{EMFFs of $\Delta^0$.}}
        \label{EMFigure0}
    \end{figure}

The flavor wave function of $\Delta^0$ implies that the contributions from the $u$ and $d$ quarks to its EMFFs would cancel in the absence of pion cloud correction.
However, this cancellation is broken by the pion cloud effect, which induces different effective coupling vertices for the $u$ and $d$ quarks, resulting in non-zero EMFFs for $\Delta^0$.
As seen in Fig.~\ref{EMFigure0}, the electric-monopole and magnetic-dipole form factors of $\Delta^0$ have the similar shape with those of the neutron. The obtained physical properties, including the electromagnetic radii, magnetic moment and electric-quadrupole moment, are consistent with the non-zero results from other studies shown in Tabs.~\ref{ChargeRadius}-\ref{EQM}.

% Table~\ref{EQM} shows the electric-quadrupole moments, which characterize the charge distribution deformations of the particle.
% The positive electric-quadrupole moment suggests that the charge distribution inside the particle has a prolate shape, and on the contrary, the negative value stands for the oblate distribution.
% As seen in Tab.~\ref{EQM},  $\Delta^+$ has an oblate charge distribution due to its negative quadrupole moment.

\begin{table}[H]
    \renewcommand\arraystretch{1.3}
    \centering
    \caption{\small{Electric charge radii of $\Delta$ isobars, compared with the results from the lattice QCD~\cite{Alexandrou:2008bn}, the chiral perturbation theory~\cite{Li:2016ezv,Geng:2009ys}, the chiral constituent quark model~\cite{Berger:2004yi}, the chiral quark model~\cite{Wagner:2000ii}, the $1/N_c$ expansion~\cite{Flores-Mendieta:2015wir,Buchmann:2002et}, and the chiral quark-soliton model~\cite{Kim_Kim_2019}.}}
    \scalebox{0.95}{\begin{tabular}{lcccccccccc}
    \toprule
    \toprule
        ${\langle r^2\rangle}_{E0}/ \text{fm}^2$ & Dressed & Point-like & \cite{Alexandrou:2008bn} & \cite{Li:2016ezv} & \cite{Geng:2009ys} & \cite{Berger:2004yi} & \cite{Wagner:2000ii} & \cite{Flores-Mendieta:2015wir} & \cite{Buchmann:2002et} & \cite{Kim_Kim_2019}\\
        \midrule
        $\Delta^{++}$ & 0.84 & 0.72 & $\cdots$ & 0.30(11) & 0.325(22) & 0.43 & 0.77 & 1.048 & 0.783 & 0.826 \\
        $\Delta^{+}$ & 0.81 & 0.72 & 0.411(28) & 0.29(10) & 0.328(21) & 0.43 & 0.77 & 1.101 & 0.783 & 0.792 \\
        $\Delta^{0}$ & $-0.06$ & 0 & $\cdots$ & $-0.02(1)$ & 0.006(1) & 0 & 0 & 0.105 & 0 & $-0.069$\\
        $\Delta^{-}$ & 0.93 & 0.72 & $\cdots$ & 0.33(11) & 0.316(23) & 0.43 & 0.77 & 0.891 & 0.783 & 0.930 \\
        \bottomrule 
        \bottomrule 
    \end{tabular}}
    \label{ChargeRadius}
\end{table}

\begin{table}[H]
    \renewcommand\arraystretch{1.3}
    \centering              
    \caption{\small{Magnetic radii of $\Delta$ isobars, compared with the results from the chiral perturbation theory~\cite{Li:2016ezv}, the chiral quark model~\cite{Wagner:2000ii}, and the chiral quark-soliton model~\cite{Kim_Kim_2019}.}}
    \begin{tabular}{lccccc}
    \toprule
    \toprule
        ${\langle r^2\rangle}_{M1}/ \text{fm}^2$ & Dressed & Point-like & \cite{Li:2016ezv} & \cite{Wagner:2000ii} & \cite{Kim_Kim_2019} \\
        \midrule
        $\Delta^{++}$ & 0.83 & 0.70 & 0.61(15) & 0.62 & 0.587\\
        $\Delta^{+}$ & 0.80 & 0.70 & 0.64(14) & 0.62 & 0.513\\
        $\Delta^{0}$ & 1.14 & 0 & 0.07(12) & 0 & 1.786\\
        $\Delta^{-}$ & 0.90 & 0.70 & 0.55(19) & 0.62 & 0.764 \\
        \bottomrule 
        \bottomrule 
    \end{tabular}
    \label{MagneticRadius}
\end{table}

\begin{table}[H]
    \renewcommand\arraystretch{1.3}
    \centering      
    \caption{\small{Magnetic moments of $\Delta$ isobars, compared with the results from PDG~\cite{ParticleDataGroup:2024cfk}, the chiral perturbation theory~\cite{Geng:2009ys}, the lattice QCD~\cite{Leinweber:1992hy}, the chiral quark model~\cite{Wagner:2000ii}, the chiral quark-soliton model~\cite{Kim_Kim_2019}, the relativistic quark model~\cite{Schlumpf:1993rm}, the QCD sum rules~\cite{Lee:1997jk}, and the $1/N_c$ expansion~\cite{Luty:1994ub}.}}
    \scalebox{0.9}{\begin{tabular}{lcccccccccc}
    \toprule
    \toprule
        $\mu/\mu_N$ & Dressed & Point-like & \cite{ParticleDataGroup:2024cfk} & \cite{Geng:2009ys} & \cite{Leinweber:1992hy} & \cite{Wagner:2000ii} & \cite{Kim_Kim_2019} & \cite{Schlumpf:1993rm} & \cite{Lee:1997jk} & \cite{Luty:1994ub}\\
        \midrule
        $\Delta^{++}$ & 5.71 & 4.86 & 6.14(51) & 6.04(13) & 4.91(61) & 6.93 & 3.65 & 4.76 & 4.13(1.30) & 5.9(4)\\
        $\Delta^{+}$ & 2.59 & 2.43 & $2.7^{+1.0}_{-1.3} \pm 1.5 \pm 3$ & 2.84(2) & 2.46(31) & 3.47 & 1.72 & 2.38 & 2.07(65) & 2.9(2)\\
        $\Delta^{0}$ & $-0.54$ & 0 & $\cdots$ & -0.36(9) & 0.00 & 0 & $-0.21$ & 0 & 0 & $\cdots$\\
        $\Delta^{-}$ & $-3.66$ & $-2.43$ & $\cdots$ & $-0.356(20)$ & $-2.46(31)$ & $-3.47$ & $-2.14$ & $-2.38$ & $2.07(65)$ & $-2.9(2)$\\
        \bottomrule 
        \bottomrule 
    \end{tabular}}
    \label{MagneticMoments}
\end{table}

 \begin{table}[H]
    \renewcommand\arraystretch{1.3}
    \centering           
    \caption{\small{Electric-quadrupole moments of $\Delta$ isobars, compared with the results from the lattice QCD~\cite{Alexandrou:2008bn}, the chiral quark model~\cite{Wagner:2000ii}, the $1/N_c$ expansion~\cite{Buchmann:2002et}, the chiral quark-soliton model~\cite{Kim_Kim_2019}, the Skyrme model~\cite{Oh:1995hn}, the non-relativistic quark model~\cite{Krivoruchenko:1991pm}, and the QCD sum rules~\cite{Azizi:2009egn}.}}
    \begin{tabular}{lccccccccc}
    \toprule
    \toprule
        $\mathcal{Q}/ \text{fm}^{2}$ & Dressed & Point-like & \cite{Alexandrou:2008bn} & \cite{Wagner:2000ii} & \cite{Buchmann:2002et} & \cite{Kim_Kim_2019} & \cite{Oh:1995hn} & \cite{Krivoruchenko:1991pm} & \cite{Azizi:2009egn}\\
        \midrule
        $\Delta^{++}$ & $-0.146$ & $-0.136$ & $\cdots$ & $-0.252$ & $-0.120$ & $-0.102$ & $-0.088$ & $-0.093$ & $-0.028(8)$\\
        $\Delta^{+}$ & $-0.070$ & $-0.068$ & $-0.019(17)$ & $-0.126$ & $-0.060$ & $-0.039$ & $-0.029$ & $-0.046$ & $-0.014(4)$\\
        $\Delta^{0}$ & 0.006 & 0 & $\cdots$ & 0 & 0 & 0.032 & 0.029 & 0 & 0\\
        $\Delta^{-}$ & 0.082 & 0.068 & $\cdots$ & 0.126 & 0.060 & 0.085 & 0.088 & 0.046 & 0.014(4)\\
        \bottomrule 
        \bottomrule 
    \end{tabular}
    \label{EQM}
\end{table}

\subsection{Gravitational form factors of $\Delta$}

\begin{figure}[H]
    \centering
    \includegraphics[width=0.46\linewidth]{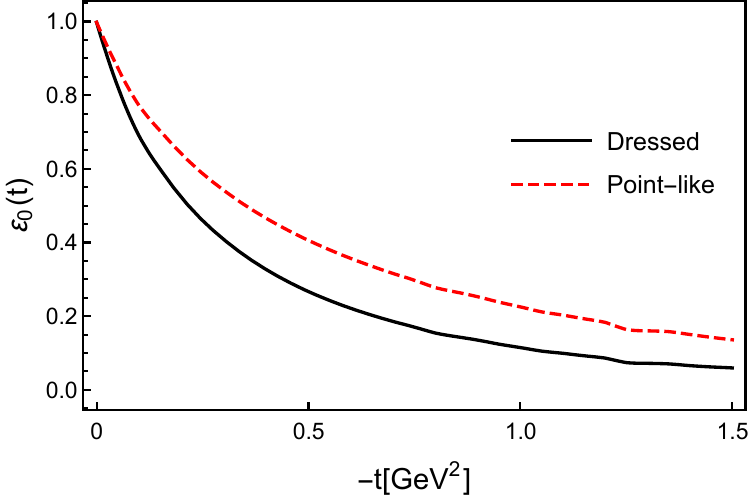} 
    \includegraphics[width=0.46\linewidth]{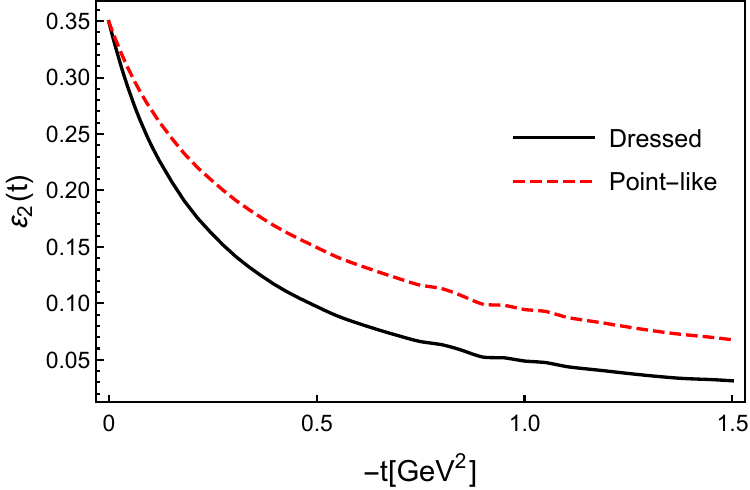} 
    \includegraphics[width=0.46\linewidth]{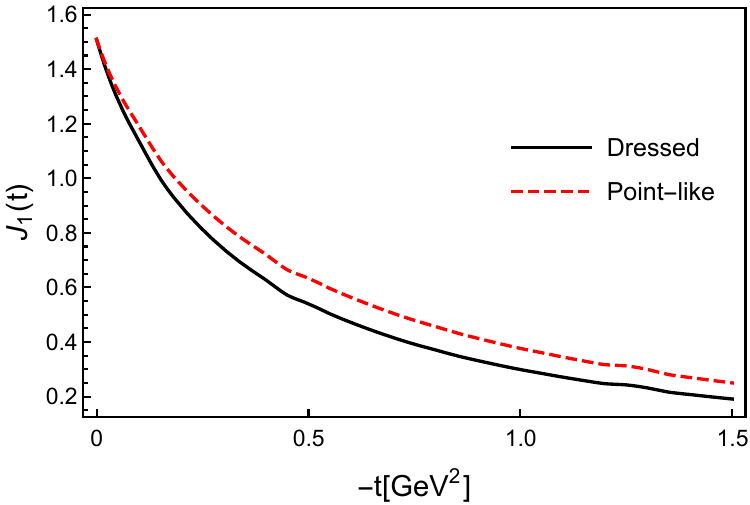}
    \includegraphics[width=0.46\linewidth]{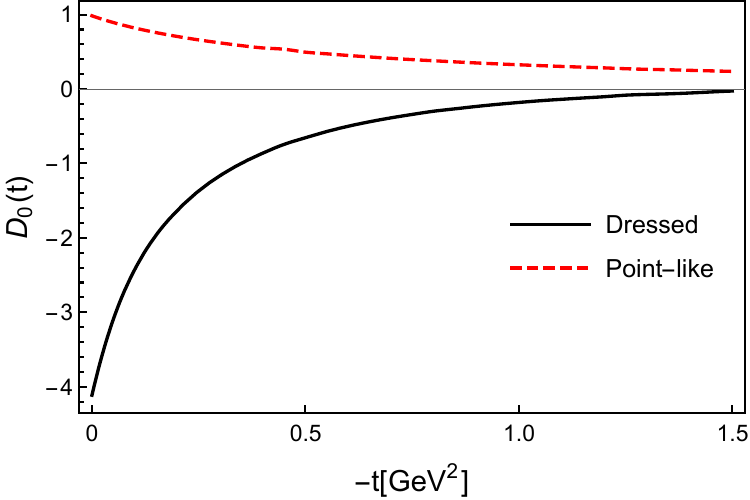}
    \caption{\small{Calculated GFFs of the $\Delta$ resonance, including $\varepsilon_0(t)$, $\varepsilon_2(t)$, $\mathcal{J}_1(t)$ and $D_0(t)$.}}
    \label{GFigure}
\end{figure}

The numerical results of the $\Delta$'s GFFs are illustrated in Fig.~\ref{GFigure}. 
Since the GFFs are independent of the charge, those of different isospin states of $\Delta$ are indistinguishable.
As discussed earlier, the momentum-dependent scalar function in Eq.~\eqref{vertexfunction2} may break the Ward-Takahashi identity~\cite{Broniowski:2008hx,Davidson:1994uv}.
Consequently, the form factors $G_E(t)$ and $\varepsilon_0(t)$ cannot be normalized exactly at the same time.
As shown in Fig.~\ref{GFigure}, when $t$ goes to zero, the results are $\varepsilon_0(0)\approx 1.00$ and $\mathcal{J}_1(0)\approx 1.51$, which are still in close agreement with the normalization condition $\varepsilon_0(0)=1$ and $\mathcal{J}_1(0)=3/2$.
% Similar with the electric-quadrupole moment, the energy-quadrupole moment derived from $\varepsilon_2(0)$ characterizes the shape of the mass distribution inside the baryon.
% As discussed before, the positive quadrupole moment represents that the mass distribution has the prolate shape.

The mass radii derived from $\varepsilon_0(t)$ are listed in Tab.~\ref{MassRadiusTable}. For the point-like results, the charge radius of $\Delta^+$ is larger than the mass radius, while the mass radius is enlarged with the pion cloud effect. This may be attributed to the neutral particle $\pi^0$, which leaves no effect on the EMFFs but could contribute to the GFFs.

The pion cloud leaves little influence to the energy and angular-momentum form factors of $\Delta$ when $t=0$. On the contrary, the well-known D-term $D_0(0)$ is sensitive to the cloud correction.
The D-term is expected to be negative to guarantee the mechanical stability of the system~\cite{Perevalova:2016dln}.
As seen in Fig.~\ref{GFigure} and Tab.~\ref{MassRadiusTable}, the obtained D-term is positive without the pion cloud effect, but the pion cloud provides a negative contribution and changes the sign of the D-term into negative.
This is consistent with our previous work and Refs.~\cite{Goeke:2007fp,Fujita:2022jus,Sugimoto:2025btn}, which focus on the nucleon form factors.
Moreover, Tab.~\ref{MassRadiusTable} shows that the mechanical radius $\langle r^2 \rangle_{\text{mech}}$ derived from $D_0(t)$ is close to the mass radius $\langle r^2 \rangle_{\text{m}}$.
\begin{table}[H]
    \renewcommand\arraystretch{1.2}
    \centering
    \caption{\small{The mass radius, mechanical radius and D-term of the $\Delta$ resonance, compared with the results from the Skyrme model~\cite{Kim:2020lrs,Perevalova:2016dln}, the QCD sum rule~\cite{Dehghan:2023ytx}, and the lattice QCD (only the gluon contribution)~\cite{Pefkou:2021fni}.}} 
    \begin{tabular}{p{2.0cm}<{\centering} p{1.8cm}<{\centering} p{1.8cm}<{\centering} 
    p{1.6cm}<{\centering} p{1.6cm}<{\centering} p{1.8cm}<{\centering} p{2.4cm}<{\centering}}
        \toprule
        \toprule
         & Dressed & Point-like & \cite{Kim:2020lrs} & \cite{Perevalova:2016dln} & \cite{Dehghan:2023ytx} & \cite{Pefkou:2021fni}\\
        \midrule
        $D_0(0)$ & $-4.11$ & 0.98 & $-3.53$ & $-3.31$ & $-2.71 (34)$ & $-1.80 (69)$\\ 
        \midrule
        $\langle r^2 \rangle_{\text{m}} /\text{fm}^2$ & 0.92 & 0.62 & 0.64 & 0.64 & $0.67 (4)$ & $0.151 (26)$\\ 
        \midrule
        $\langle r^2 \rangle_{\text{mech}}/\text{fm}^2$ & 0.91 & 0.29 & 0.85 & $\cdots$  & $\cdots$ & $0.355(113)$\\ 
        \bottomrule
        \bottomrule
    \end{tabular}
    \label{MassRadiusTable}  
\end{table} 

\begin{figure}[H]
    \centering
    \includegraphics[width=0.46\linewidth]{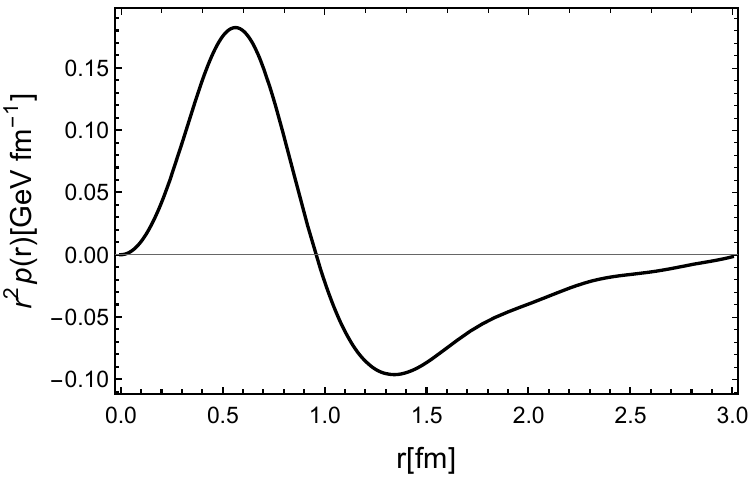}        
    \includegraphics[width=0.46\linewidth]{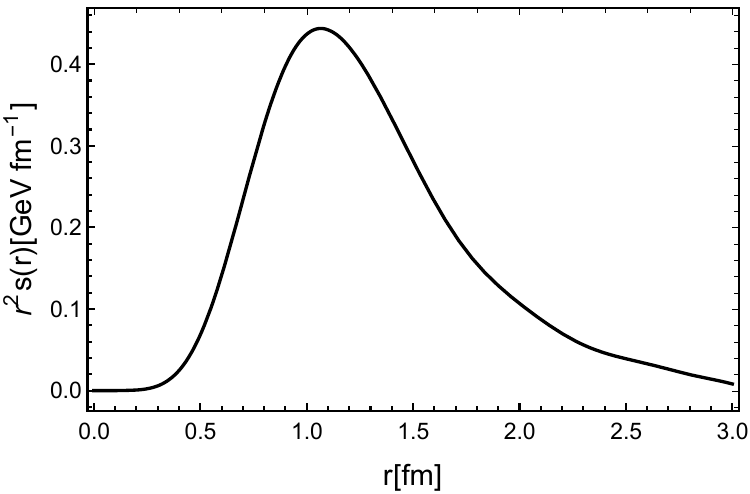}
    \caption{\small{The pressure and shear force distributions inside $\Delta$.}}
    \label{PSF}
\end{figure}

Through the Fourier transformation, one accesses the mechanical properties in the $r$-space. To localize the particle~\cite{Epelbaum:2022fjc,Diehl:2002he,Freese:2021mzg} and guarantee the convergence of the Fourier transformation, a Gaussian-like wave packet $e^{t/\lambda^2}$~\cite{Ishikawa:2017iym} is attached to the GFFs during the integrals, where $1/\lambda$ correlates with the size of the hadron. Here we choose $\lambda=0.92 \,\text{GeV}$ to ensure the zero crossing point $r_0$ of the pressure distribution satisfies $r_0 \sim \sqrt{\langle r^2 \rangle_{\text{m}}}$ (seen in Fig.~\ref{PSF}).

The pressure and shear force distributions derived from $D_0(t)$ are illustrated with black solid curves in Fig.~\ref{PSF}.
As discussed earlier, the pion cloud changes the sign of the D-term to ensure the mechanical stability inside the particle.
As a consequence, the pressure and shear force obtained with and without the pion cloud have different signs as well. Here we only present the dressed results in Fig.~\ref{PSF}, and the repulsive (positive) and binding (negative) pressures are separated by the zero crossing point $r_0 \approx \sqrt{\langle r^2 \rangle_{\text{m}}}$ as discussed above, and the peak and the valley of the pressure are $r\approx0.56\, \text{fm}$ and $r\approx1.34\, \text{fm}$.
Moreover, the pressure distribution satisfies the Von Laue condition $\int^\infty_0 dr r^2 p_0(r)=0$, which is necessary to maintain the mechanical stability of the system.

\begin{figure}[H]
    \centering
    \includegraphics[width=0.46\linewidth]{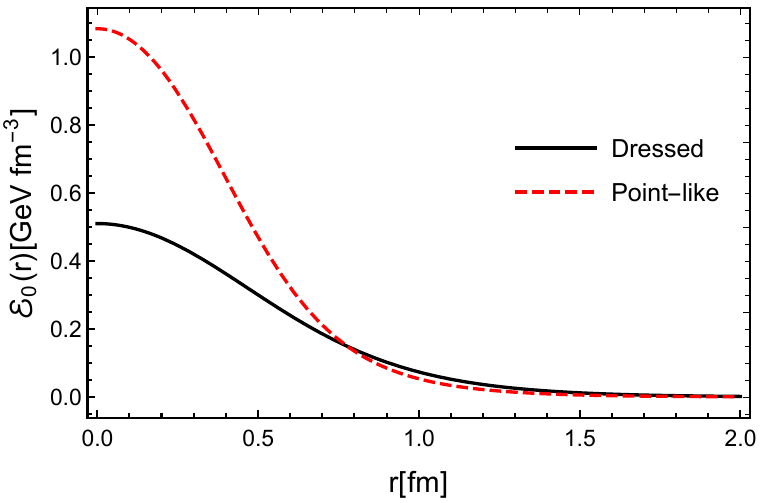}        
    \includegraphics[width=0.46\linewidth]{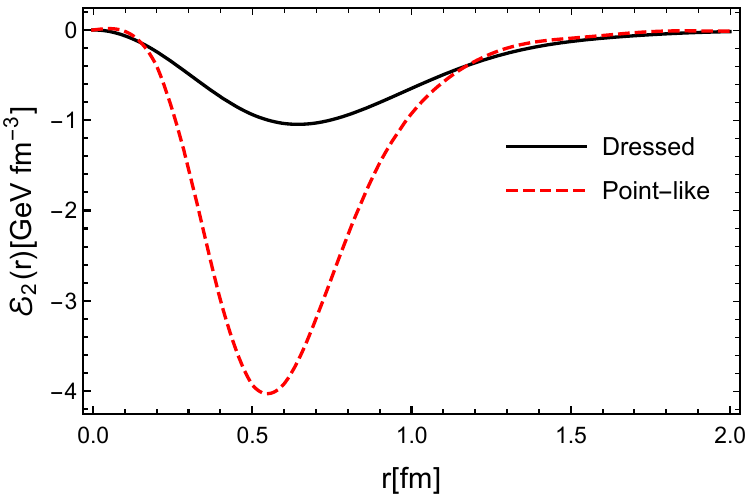}   
    \includegraphics[width=0.46\linewidth]{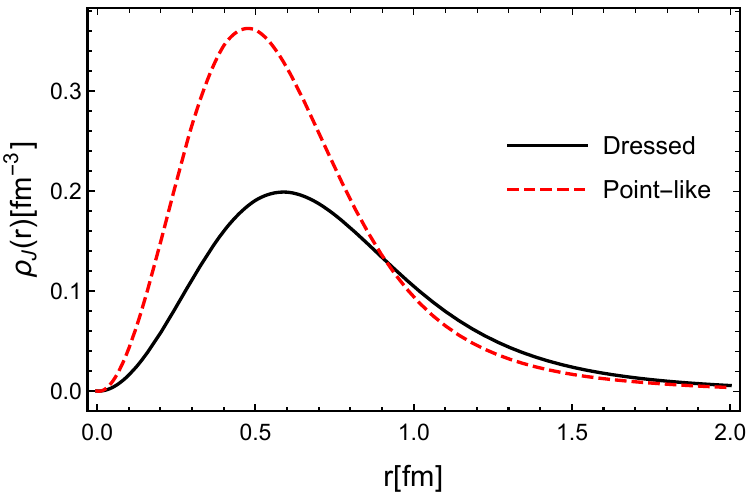}
    \caption{\small{The energy-monopole, energy-quadrupole and angular momentum densities of $\Delta$.}}
    \label{EAMDens}
\end{figure}
    
The energy and angular momentum densities derived from ${\varepsilon}_{0(2)}(t)$ and $\mathcal{J}_{1}(t)$ are illustrated in Fig.~\ref{EAMDens}. Comparing the two curves, the pion cloud depresses the densities at small $r$, while the densities with the pion cloud effect are slightly larger when approximately $r > \sqrt{\langle r^2\rangle_{\text{m}}}$. For the dressed angular momentum density, the maximum is at $r\approx0.59\, \text{fm}$, and for the result without the pion cloud, the maximum is at $r\approx0.48\, \text{fm}$. It suggests that the pion cloud could distract the mass and angular momentum from the origin, which is consistent with the intuitive expectation.

\subsection{Electromagnetic form factors of $N-\Delta$ transition}

\begin{figure}[htbp]
    \centering
    \includegraphics[width=0.46\linewidth]{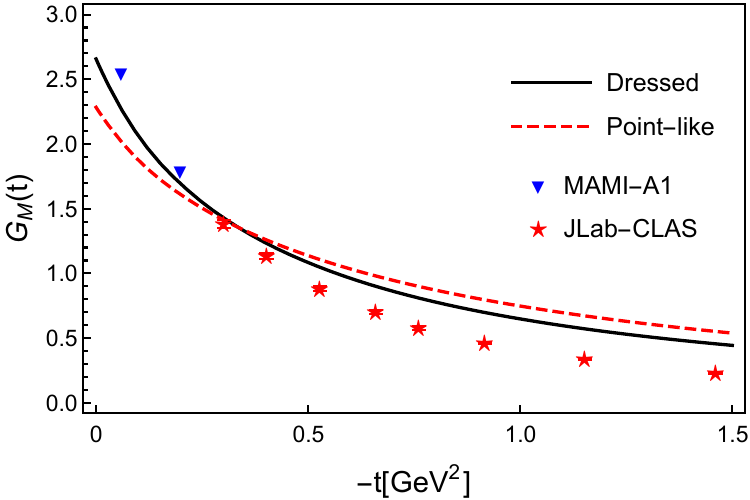}
    \includegraphics[width=0.46\linewidth]{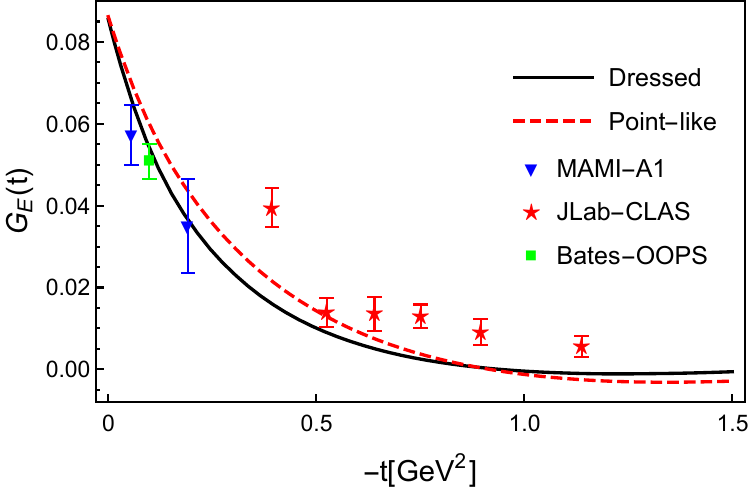}        
    \includegraphics[width=0.47\linewidth]{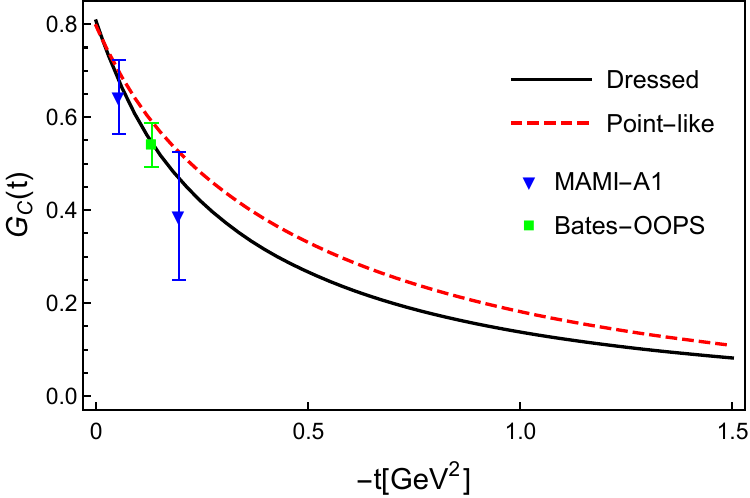}
    \caption{The electromagnetic $N-\Delta$ transition form factors compared with the experimental data from JLab/CLAS (red stars)~\cite{CLAS:2009ces,CLAS:2001cbm}, MAMI (blue triangles)~\cite{A1:2008ocu}, and MIT-bates (green squares)~\cite{Mertz:1999hp}.}
\label{NDFigure}
\end{figure}

The results of the transition form factors are illustrated in Fig.~\ref{NDFigure}, where the coloured points stand for the experimental results~\cite{CLAS:2009ces,CLAS:2001cbm,A1:2008ocu,Mertz:1999hp}.
Our numerical results show their good agreement with the experimental data in Fig.~\ref{NDFigure}, and the obtained results with $t=0$ are consistent with other experimental and theoretical studies shown in Tab.~\ref{NDTable}.
The magnetic-dipole form factor is smaller than the experimental results because our model is still not accurate enough.
As seen in Fig.~\ref{f-emff}, only three diagrams are taken into consideration. Some other physical ingredients, such as the quark-diquark exchange and the so-called seagull contribution are neglected during the calculation, which may lead to the discrepancy.
Besides, the introduction of the scalar function without solving the Bethe-Salpeter equation also brings uncertainties in our results.    
    
Comparing with the obtained EMFFs of $\Delta^+$, the pion cloud effect on the transition magnetic-dipole form factor is stronger. The magnetic-dipole moment and electric-quadrupole moment are derived from the form factors as     
\begin{equation}
\label{NDMoment}
    \mu_{N -\Delta}=3.05 \mu_N,\quad \quad  \mathcal{Q}_{N-\Delta}=-0.062\,\text{fm}^2.
\end{equation}

In the quark-model picture, the spin flip of the $s$-wave quark can only lead to the magnetic-dipole transition form factors.
Therefore, the non-zero quadrupole transition form factors indicate that there are possible $d$-wave components within the nucleon and $\Delta$ wave functions~\cite{Isgur:1981yz,Capstick:1989ck}, which will lead to the deformations of the charge distributions inside the baryons.
It should be mentioned that the deformation inside the nucleon is known as the intrinsic deformation, which cannot be measured experimentally due to its spin average~\cite{Buchmann:2001gj}. The value of the intrinsic quadrupole moment is model-dependent.
In our current study, we can only infer that the intrinsic quadrupole moment of the nucleon is non-zero, and the exact value of the quadrupole moment needs further studies in the future.

As seen in Fig.~\ref{NDFigure}, the results with the pion cloud fit the experimental data better.
Compared with the point-like results, $G_E(t)$ and $G_C(t)$ fall off faster and the magnetic-dipole moment is raised about $16\%$.
Different from the dipole moment, the pion cloud leaves little effect on the quadrupole moment.
However, the studies such as Ref.~\cite{Pascalutsa:2005ts} (the chiral EFT) and Ref.~\cite{Lu:1996rj} (the chiral bag model) argue that the pion cloud contribution takes a dominant position in the quadrupole form factors.
This may be attributed to the different interpretation of the pion cloud.
In our model, the pion cloud are coupling with the quarks inside the baryon instead of the baryon itself, which may lead to different 
Moreover, the pion exchange among the quarks is not taken into consideration, and
this may lead to the limited contributions.
    
When $t$ goes to zero, the helicity amplitudes reduce to the photon-couplings as
\begin{equation}
\label{NDHA}
    A_{1/2}=-125 \times 10^{-3}\,\text{GeV}^{-1/2}, \quad A_{3/2}=-233 \times 10^{-3}\,\text{GeV}^{-1/2}, \quad S_{1/2}=-12.5 \times 10^{-3}\,\text{GeV}^{-1/2}.
\end{equation}    
The results are compared with point-like results and the results from the experiments and other studies in Tab.~\ref{NDTable}.
The pion cloud enlarges the magnitudes of $A_{1/2}$ and $A_{3/2}$ since $G_M(0)$ is raised by the pion cloud correction.
Compared with the experimental data, our results are underestimated by about $10\%$ but still qualitatively consistent with it. 
    
\begin{table}[H]
    \renewcommand\arraystretch{1.3}
    \centering           
    \caption{\small{Physical properties derived from the transition form factors, including the helicity amplitudes (in unit of $10^{-3}\,\text{GeV}^{-1/2}$), $R_{EM}$ and $R_{SM}$. The results are compared with the experimental data~\cite{ParticleDataGroup:2024cfk}, and theoretical studies including the chiral bag model~\cite{Fiolhais:1996bp,Lu:1996rj,Bermuth:1988ms}, the quark-diquark model~\cite{Keiner:1996ch}, the relativistic quark model~\cite{Dong:2001js}, the chiral EFT~\cite{Pascalutsa:2005ts}, the Dyson-Schwinger approach~\cite{Eichmann:2011aa,Segovia:2013rca} and the chiral quark-soliton model~\cite{Kim:2020lgp}.}}
    \begin{tabular}{lcccc}
    \toprule
    \toprule
         & $A_{1/2} $ & $A_{3/2}$  & $R_{EM}$ & $R_{SM}$\\
        \midrule
        Dressed & $-125$ & $-233$ & $-3.21\%$ & $-2.90\%$ \\
        Point-like & $-106$ & $-199$ & $-3.77\%$ & $-3.67\%$ \\
        PDG~\cite{ParticleDataGroup:2024cfk}  & $-135^{+6}_{-7}$ & $-255 ( 7)$ & $-2.5 (3)\%$ & $\cdots$  \\
        CBM~\cite{Fiolhais:1996bp} & $-107$ & $-199$ & $-1.8 \%$ & $-2.3 \%$   \\
        CBM~\cite{Lu:1996rj} & $-128$ & $-222$ & $-2.9 \%$ & $\cdots$   \\
        CBM~\cite{Bermuth:1988ms} & $\cdots$ & $-194$ & $-1.8 \%$ & $\cdots$   \\
        QDM~\cite{Keiner:1996ch} & $\cdots$ & $\cdots$ & $-3.4\%$ & $2.1\%$   \\
        RQM~\cite{Dong:2001js} & $-147$ & $-277$ & $-2.0\%$ & $\cdots$   \\
        $\chi$EFT~\cite{Pascalutsa:2005ts}  & $\cdots$ & $\cdots$ &  &   \\
        DSA~\cite{Eichmann:2011aa}  & $\cdots$ & $\cdots$ & $-2.3(3)\%$ & $-2.2(6)\%$   \\  
        DSA~\cite{Segovia:2013rca}  & $\cdots$ & $\cdots$ & $-6.2\%$ & $-2.4\%$   \\ 
        $\chi$QSM~\cite{Kim:2020lgp}  & $\cdots$ & $\cdots$ & $-1.8\%$ & $-2.3\%$   \\
        \bottomrule 
        \bottomrule 
    \end{tabular}
    \label{NDTable}
\end{table}

\section{Summary And Discussion}

The form factors of $\Delta$ isobars and the electromagnetic $N-\Delta$ transition are studied simultaneously with a relativistic covariant quark-diquark approach and with the pion cloud correction.
The quark-diquark approach could simplify the particle structure into a two-body problem, and an additional scalar function is employed to simulate the bound state between the quark and the diquark.
In this work, the quark is considered as a dressed quark coupling with the pion cloud. Due to the pion cloud effect, the interaction vertex is modified as well.
 
The obtained EMFFs in this work are consistent with those from other studies except for that the charge radius may be overestimated.
%The negative electromagnetic-quadrupole moment suggests that the charge deformation of $\Delta^+$ has the oblate shape.
Unlike our previous work about the proton~\cite{Wang:2024abv}, the pion cloud effect on the $\Delta^+$ EMFFs is nearly negligible.
Different from $\Delta^+$, the pion cloud raises the magnitudes of the magnetic moment and electric-quadrupole moment of $\Delta^-$ significantly, and provides a non-zero contribution to the EMFFs of $\Delta^0$. It results from the different pion cloud correction between the $u$ and $d$ quarks, which breaks the isospin symmetry among the $\Delta$ isobars. 
    
With regard to the GFFs, although the introduced momentum-dependent scalar function may break the Ward-Takahashi identity, the energy-monopole and angular momentum-dipole form factors could almost satisfy the normalization condition well.
% Different from the charge distribution, the positive energy-quadrupole moment represents that the mass distribution has the prolate shape.
The pion cloud provides a negative contribution to the D-term, which plays an indispensable role to guarantee the mechanical stability of the system. 
Similar with the charge radius, the pion cloud raises the mass and mechanical radii of $\Delta$.
Compared with the charge and magnetic radii, the mass radius is larger due to the pion cloud effect from the neutral $\pi^0$. 
Furthermore, the mechanical properties in the coordinate space, including the energy density, angular momentum density, and the inner force distributions are derived from the GFFs. The pion cloud effect distracts the mass and angular momentum from the origin, and reverses the signs of the pressure and shear force.

The EMFFs of the $N-\Delta$ transition are in good agreement with the experimental results, except for that the magnetic-dipole moment is underestimated by about $10\%$.
The non-zero quadrupole form factors suggest that there are intrinsic deformations inside the nucleon and $\Delta$.
The pion cloud effect is specially discussed in this work. It enlarges the magnetic-dipole form factor about $16\%$ and causes the quadrupole form factors to fall off faster in the small $|t|$.
It should be mentioned that we did not study the GFFs of the $N-\Delta$ transition in this work. 
Since the GFFs contributed by $u$ and $d$ quarks are indistinguishable in our model, the transition GFFs vanish naturally.
More details about the transition GFFs can be found in Refs.~\cite{Kim:2022bwn,Alharazin:2023zzc}. 

The aim of our future study is the GPDs of the $N-\Delta$ transition, which could be accessed experimentally through the hard exclusive pion production~\cite{CLAS:2023akb} and the deeply virtual Compton scattering~\cite{PhysRevD.108.034021} processes.
The EMFFs can be extracted through the GPDs, and hence results in this work can be employed in our future study to draw a comparison.
% It should be mentioned that in this paper the form factors are evaluated in a simple model, which may lead to discrepancies. 
Furthermore, it is expected that the model will be improved in future studies, and more accurate results may be obtained.

\appendix

\renewcommand\thesection{Appendix}
\section{Flavour Wave Functions of the Nucleon And $\Delta$}\label{appendix}

In the quark-diquark approach, the flavour wave functions of the nucleon and $\Delta$ can be expressed as~\cite{Lichtenberg_Tassie_Keleman_1968,Ma:2002ir}

\begin{equation}
\begin{aligned}
    &\ket{p}=\cos{\theta}|u(ud)_s\rangle+\sin{\theta}\left[\sqrt{\frac{2}{3}}|d (uu)_a\rangle-\sqrt{\frac{1}{3}}|u (ud)_a\rangle\right],\\
    &\ket{n}=\cos{\theta}|d(ud)_s\rangle+\sin{\theta}\left[-\sqrt{\frac{2}{3}}|u (dd)_a\rangle+\sqrt{\frac{1}{3}}|d(ud)_a\rangle\right],\\
    &\ket{\Delta^{++}} =\ket{u(uu)_a}, \\
    &\ket{\Delta^{+}} 
    =\sqrt{\frac{2}{3}}\ket{u(ud)_a}+\sqrt{\frac{1}{3}}\ket{d(uu)_a}, \\
    &\ket{\Delta^{0}} 
    =\sqrt{\frac{2}{3}}\ket{d(ud)_a}+\sqrt{\frac{1}{3}}\ket{u(dd)_a}, \\
    &\ket{\Delta^{-}} =\ket{d(dd)_a},
\end{aligned}
\tag{A1}
\end{equation}
where $\theta$ is a mixing angle characterizing the proportion between the scalar diquark and the axialvector diquark inside the nucleon.

\section*{Acknowledgements}
We are grateful to S. Kumano for the constructive discussions. This work is supported by the National Natural Science Foundation of China under Grants No. 12375142 and No. 12447121, and the Gansu Province Postdoctor Foundation.

\bibliographystyle{unsrt}
\bibliography{references}

\end{document}